\documentclass[11pt,twoside]{article}
\usepackage{./asp2014}

\aspSuppressVolSlug
\resetcounters

\begin{document}

\title{Investigating Opacity Modifications and Reaction Rate Uncertainties to Resolve the Cepheid Mass Discrepancy}
\author{Joyce A. Guzik,$^1$ Ebraheem Farag,$^2$ Jakub Ostrowski,$^3$ Nancy R. Evans,$^4$
Hilding Neilson,$^5$ Sofia Moschou,$^4$ and Jeremy J. Drake$^4$
\affil{$^1$Los Alamos National Laboratory, Los Alamos, NM USA; \email{joy@lanl.gov}}
\affil{$^2$Arizona State U. Tempe, AZ USA; \email{ekfarag@asu.edu}}
\affil{$^3$Pedagogical U. of Cracow, Krak\'{o}w Poland; \email{jakub.ostrowski@up.krakow.pl}}
\affil{$^4$Harvard Smithsonian Center for Astrophysics, Cambridge, MA USA; \email{nevans@cfa.harvard.edu, sofia.moschou@cfa.harvard.edu,jdrake@cfa.harvard.edu}}
\affil{$^5$U. Toronto, Toronto, Ontario Canada; \email{neilson@astro.utoronto.ca}}}

\paperauthor{Sample~Author1}{Author1Email@email.edu}{ORCID_Or_Blank}{Author1 Institution}{Author1 Department}{City}{State/Province}{Postal Code}{Country}
\paperauthor{Sample~Author2}{Author2Email@email.edu}{ORCID_Or_Blank}{Author2 Institution}{Author2 Department}{City}{State/Province}{Postal Code}{Country}
\paperauthor{Sample~Author3}{Author3Email@email.edu}{ORCID_Or_Blank}{Author3 Institution}{Author3 Department}{City}{State/Province}{Postal Code}{Country}
\begin{abstract}
Cepheid masses derived from pulsations or binary dynamics are generally lower than those derived from stellar evolution models.   Recent efforts have been dedicated to investigating the effects of abundances, mass loss, rotation, convection and overshooting prescriptions for modifying the evolution tracks to reduce or remove this Cepheid mass discrepancy.  While these approaches are promising, either alone or in combination, more work is required to distinguish between possible solutions.  Here we investigate nuclear reaction rate and opacity modifications on Cepheid evolution using the MESA code.  We discuss the effects of opacity increases at envelope temperatures of 200,000-400,000 K proposed to explain the pulsation properties of main-sequence $\beta$ Cep/SPB variables which will evolve into Cepheids.  We make use of the RSP nonlinear radial pulsation modeling capability in MESA to calculate periods and radial velocity amplitudes of Galactic Cepheids V1334 Cyg, Polaris, and $\delta$ Cep.
\end{abstract}

\section{Introduction}

Interferometric observations of Cepheid binary dynamics are being used to place more stringent constraints on masses of individual Cepheids \citep[see, e.g.,][]{2018ApJ...863..187E}.  Cepheid masses have also been derived using envelope pulsation models \citep[see, e.g.,][]{2005ApJ...629.1021C}.  When these Cepheids are plotted on the Hertzsprung-Russell (H-R) diagram, their observed luminosities are higher than those of standard stellar evolution model tracks of the same mass.  In other words, the evolution models need to be 10-30\% more massive to reach the luminosities of the observed Cepheids, with the magnitude of the difference increasing toward lower mass/shorter period Cepheids.  The evolution model luminosities and the extent of 'blue loops', where stars evolve into the Cepheid pulsation instability region during their core helium-burning phase, are affected by many factors, including rotation \citep{2014A&A...564A.100A,2018A&A...616A.112S}, mass loss \citep{2006MmSAI..77..207B, 2011A&A...529L...9N}, convective overshooting \citep{2011A&A...529L...9N}, and nuclear reaction rates \citep{2010A&A...520A..41M}.  To date, no single modification of physical inputs has resolved definitively this Cepheid mass discrepancy.

Motivated by this mass discrepancy, we explore Cepheid evolution and pulsation models using the Modules for Experiments in Astrophysics (MESA) open-source stellar evolution code\footnote{mesa.sourceforge.net. See `getting started' tutorial and example ``star/test\_suite/5M\_cepheid\_blue\_loop'' for Cepheid evolution, and ``star/test\_suite/rsp\_Cepheid'' for RSP model example.} \citep{2019ApJS..243...10P}, version r12115.  In particular, we consider increases in the triple-$\alpha$ and $^{12}$C($\alpha$,$\gamma$)$^{16}$O reaction rates, and opacity enhancements that have been proposed to explain the pulsation frequencies of main-sequence hybrid $\beta$ Cepheid/Slowly-Pulsating B (SPB) stars that will evolve to become Cepheids.  We also use the new radial stellar pulsation (RSP) capability in MESA to model the nonlinear radial envelope pulsations of three Galactic Cepheids: $\delta$ Cep, Polaris, and V1334 Cyg.  

\section{Effects of reaction rate and opacity increases on Cepheid evolution}
\label{evolution}

Figure \ref{reaction_rates} compares evolution models with nominal MESA nuclear reaction rates, and with triple-$\alpha$ and $^{12}$C($\alpha$,$\gamma$)$^{16}$O reaction rates, important during helium burning, multiplied by a factor of three.  These models use AGSS09 \citep{2009ARA&A..47..481A} abundance mixture, helium mass fraction Y=0.28, metallicity Z=0.02, OPAL \citep{1996ApJ...464..943I} opacities, a standard MESA convective overshoot treatment, and do not include rotation or mass loss.  While these models with higher reaction rates do show an increase in luminosity ($\sim$0.2 dex) and extent of the blue loops, the effect is not large enough by itself to solve the mass discrepancy; for a 4.5 M$_{\odot}$ model, a luminosity increase of $\sim$0.5 dex is needed. Furthermore, reaction rates are assessed to be uncertain by less than a factor of 1.5 \citep{2018ApJS..234...19F}.

\articlefigure[width=.75\textwidth]{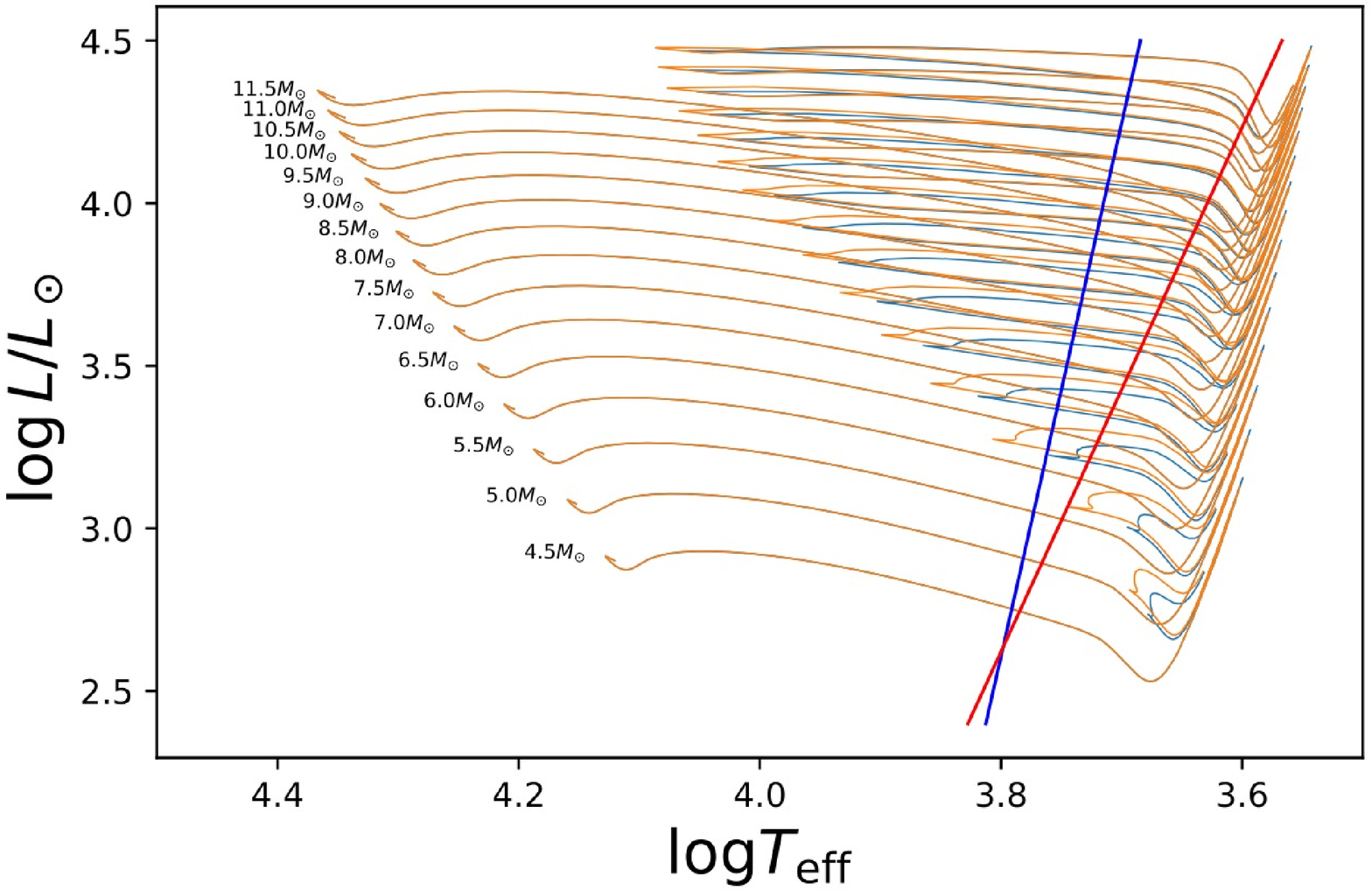}{reaction_rates}{Post-main sequence MESA evolution tracks for massive stars with unmodified reaction rates (blue), and with triple-$\alpha$ and $^{12}$C($\alpha$,$\gamma$)$^{16}$O multiplied by a factor of three (orange). The red and blue vertical lines mark the boundaries of the Cepheid instability strip for Z=0.02 from \cite{2000ApJ...529..293B}.}

Opacity increases have been proposed to explain the pulsation frequency content of main-sequence hybrid $\beta$ Cep/SPB pulsators, e.g., $\nu$ Eri \citep{2017MNRAS.466.2284D}.  We applied the same temperature-dependent multipliers for OPAL opacities used for $\nu$ Eri to Cepheid evolution models.  These multipliers are a pair of Gaussians centered around log $T$ = 5.3 and 5.46, with amplitudes of 1.5 and 2.5, respectively, and mainly affect the `Z-bump' region that causes the $\kappa$-effect driving of main-sequence B-type star pulsations.  These opacity enhancements turned out to affect Cepheid evolution tracks negligibly, only slightly decreasing the extent of the blue loop tip.  

\section{MESA Radial Stellar Pulsation (RSP) Models}
We next applied the MESA RSP capability \citep[see also][]{2008AcA....58..193S}, including a time-dependent convection treatment, to model Cepheid  envelope pulsations.  The RSP code does not allow (as yet) the import of an envelope structure and composition profile directly from an evolution model; however, envelope composition is changed insignificantly during evolution before core helium exhaustion, especially without mass loss.  The RSP model input requires mass (M), luminosity (L), effective temperature (T$_{\rm eff}$), Y and Z.  Using the nominal settings, a 150-zone envelope model is built from the stellar surface down to a temperature of 2 million K.  The linear periods and growth rates for fundamental (F), 1st overtone (OT), and 2nd OT modes are calculated including the time-dependent convection treatment.

We started with inputs based on observed parameters for Galactic Cepheids from the literature, and then varied the input parameters, mainly T$_{\rm eff}$, but also L or M if needed to identify a model calculated to have the observed pulsation period and positive linear growth rate in the desired pulsation mode.  The model's pulsation is then initialized in this mode with radial velocity amplitude 0.1 km/s.  It may be necessary to run several thousands of pulsation periods, or tens of millions of timesteps, until the model converges to a limiting radial velocity amplitude.  We applied this procedure to models of $\delta$ Cep, with estimated mass 5-5.3 M$_{\odot}$ \citep{2015ApJ...804..144A}, as well as to models of Polaris and V1334 Cyg, with dynamical masses constrained using the orbit of a binary companion of 3.45 $\pm$ 0.75 M$_{\odot}$ and 4.228 $\pm$ 0.133 M$_{\odot}$, respectively \citep{2018ApJ...867..121G}.  These models use Y=0.27, and Z=0.015 (assumed near solar) for V1334 Cyg and $\delta$ Cep, and Z=0.016, slightly above solar, for Polaris \citep{2001A&A...373..159C}. We also calculated models for each Cepheid using this same procedure including the temperature-dependent opacity multiplier discussed in Section \ref{evolution}.

\begin{figure}[h!]
\begin{center}
\includegraphics[width=8.5cm]{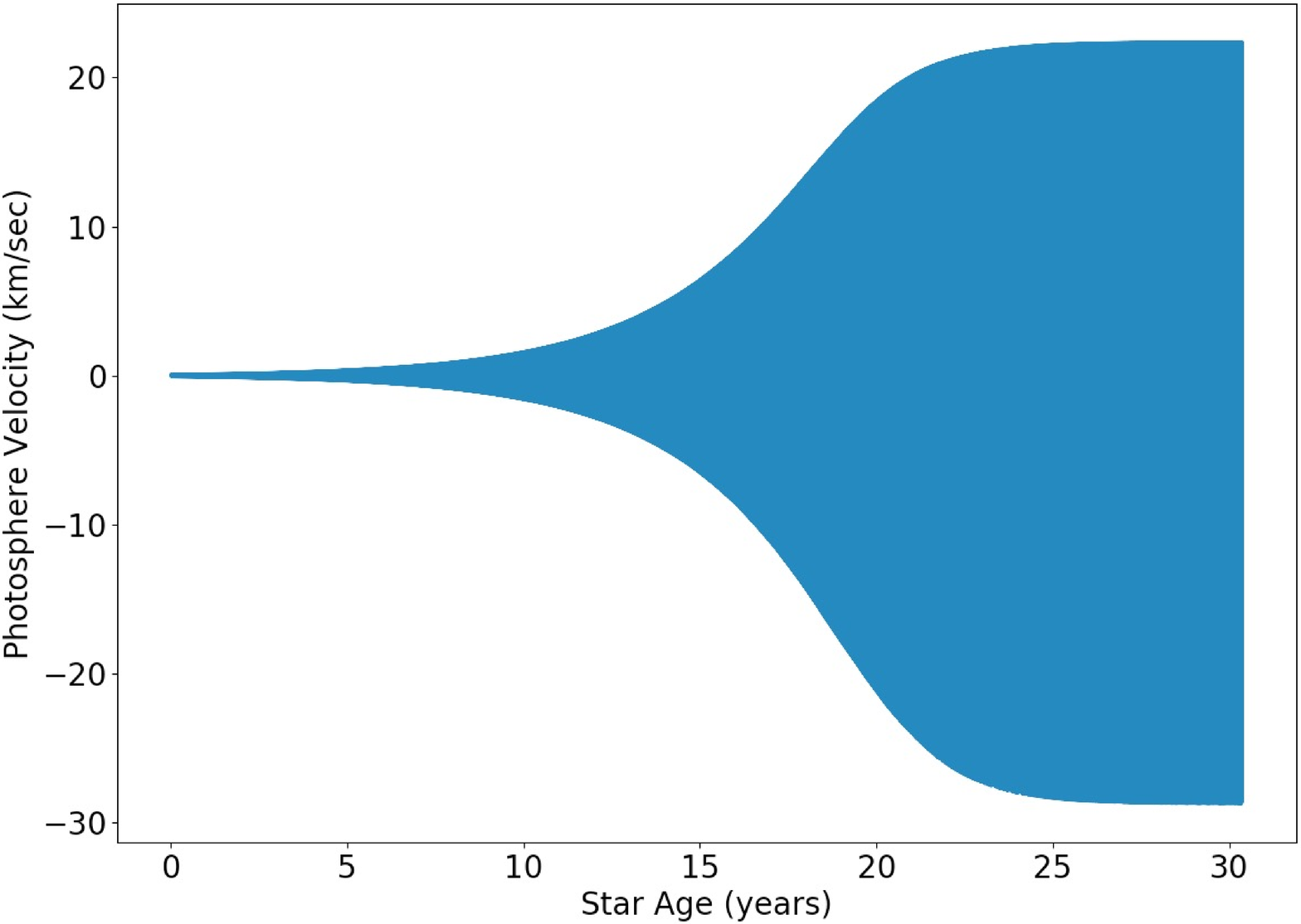}
\includegraphics[width=8.5cm]{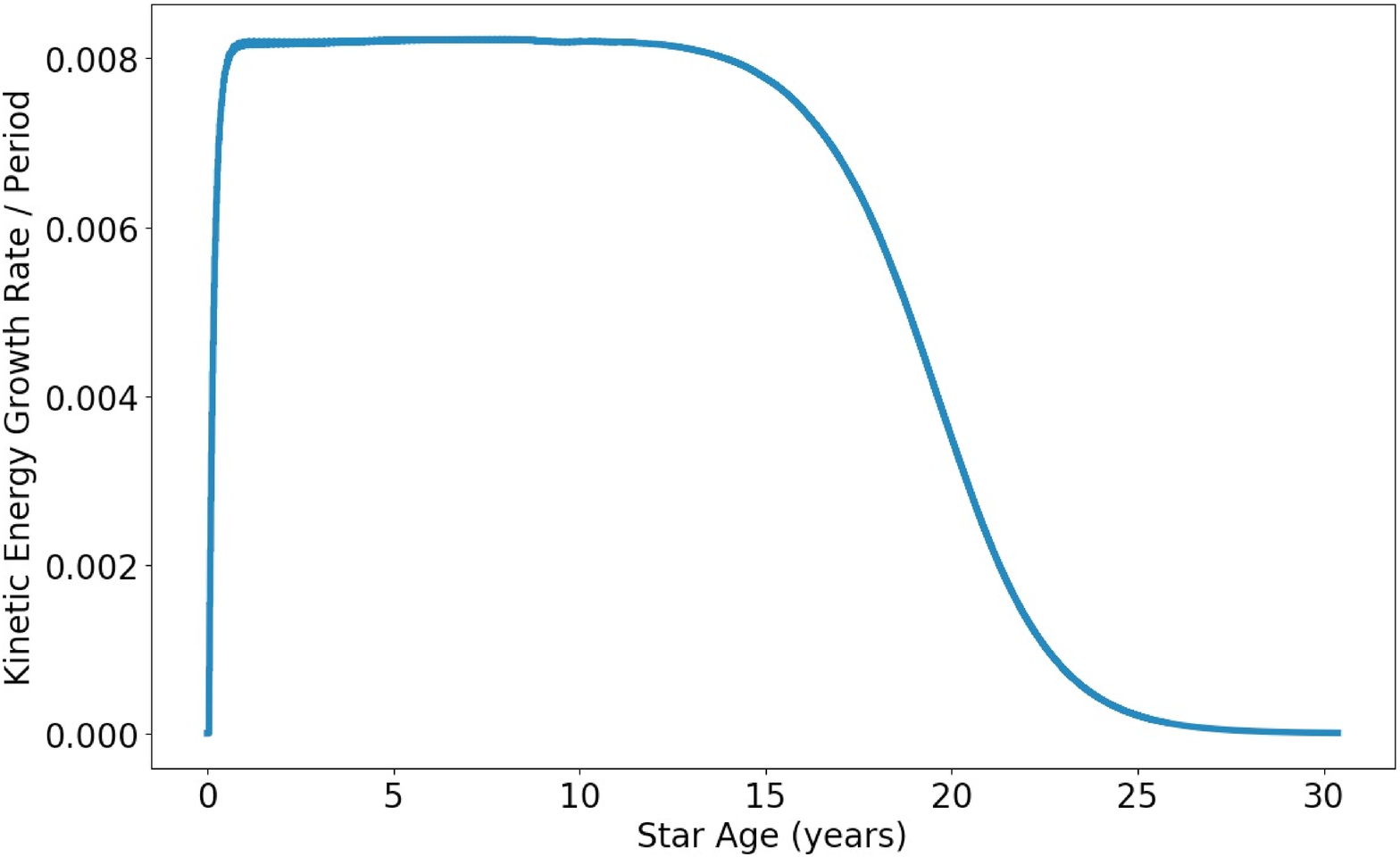}
\includegraphics[width=8.5cm]{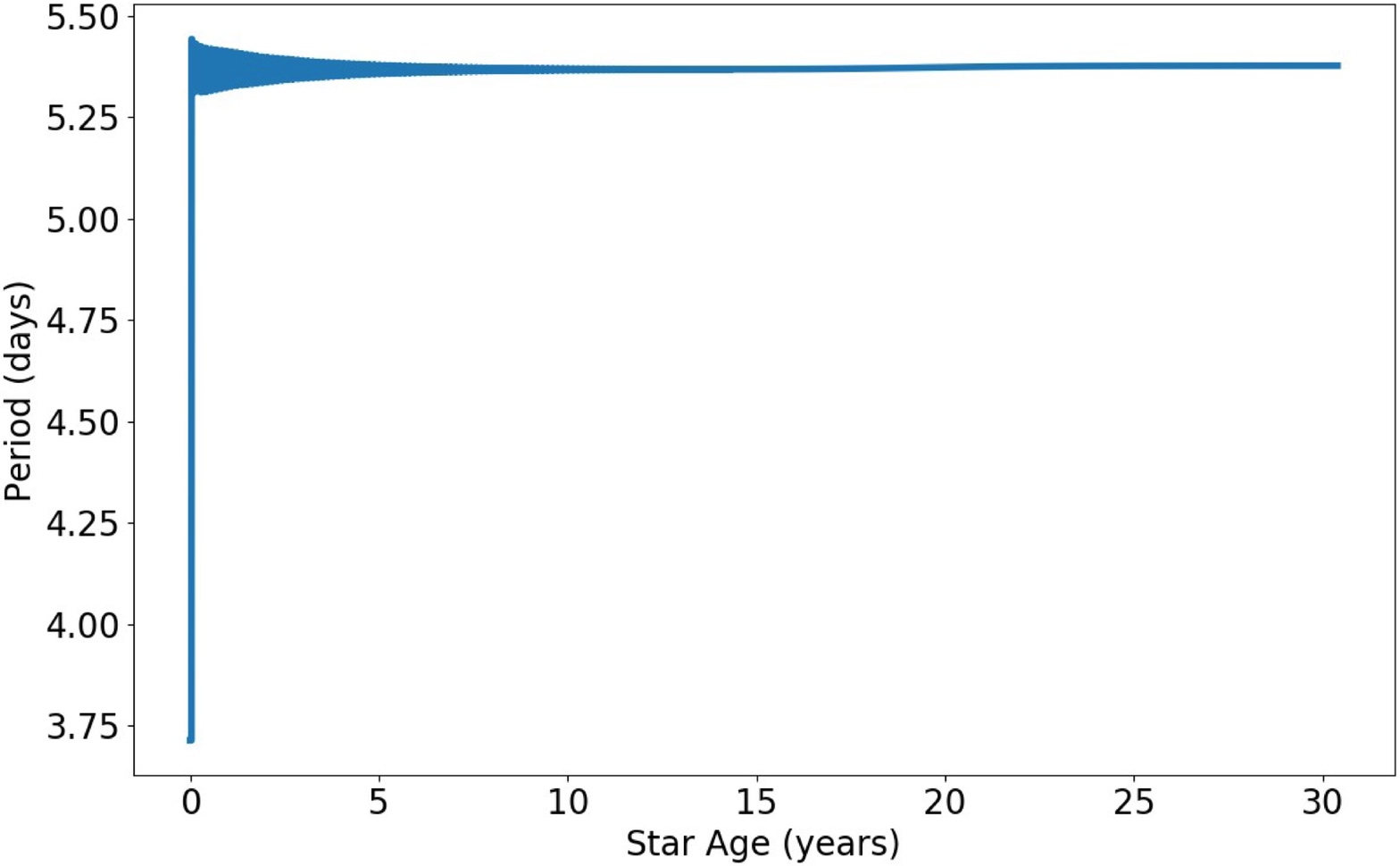}
\end{center}
\caption{Photospheric radial velocity in km/sec (top), kinetic energy growth rate per period (center), and pulsation period in days (bottom) vs. stellar age in years after the start of the hydrodynamic simulation for $\delta$ Cep RSP model without opacity enhancement.}\label{delta_Cep}
\end{figure}

Table \ref{table} summarizes the RSP model results.  Note that the Polaris model mass needed to be increased to 5.93 M$_{\odot}$ to find a model with a pulsation period in agreement with the observed value.  Figure \ref{delta_Cep} shows results for the $\delta$ Cep model without the opacity multiplier.  For the default choices of viscous dissipation in the nonlinear hydrodynamics simulation, the model reaches a limiting radial velocity amplitude 22 km/s, not far from the observed amplitude of 25 km/s \citep{2005ApJS..156..227B}.  The observed radial velocity amplitudes of Polaris and V1334 Cyg are a few km/s \citep{2008AJ....135.2240L} and about 5 km/s \citep{2018ApJ...867..121G}, respectively.  In general, the radial velocity amplitudes of the models without the opacity multiplier are in good agreement with the observed values, while the opacity multiplier causes a significant decrease in limiting amplitude. It is interesting that the V1334 Cyg model without the opacity multiplier has positive growth rates for both the F and 1st OT modes; even though the observed 1st OT mode has the highest linear growth rate and is initialized in this mode, the model switches modes after many periods to pulsate in the fundamental mode.  Figure \ref{HRD_3models} shows the observed and model L and T$_{\rm eff}$ on the H-R diagram.  Note that the V1334 Cyg and Polaris models and observations lie well above the MESA 5 M$_{\odot}$ evolution track (without rotation or mass loss), illustrating the mass discrepancy for these two stars that have dynamical mass determinations.

\clearpage

\articlefigure[width=.58\textwidth]{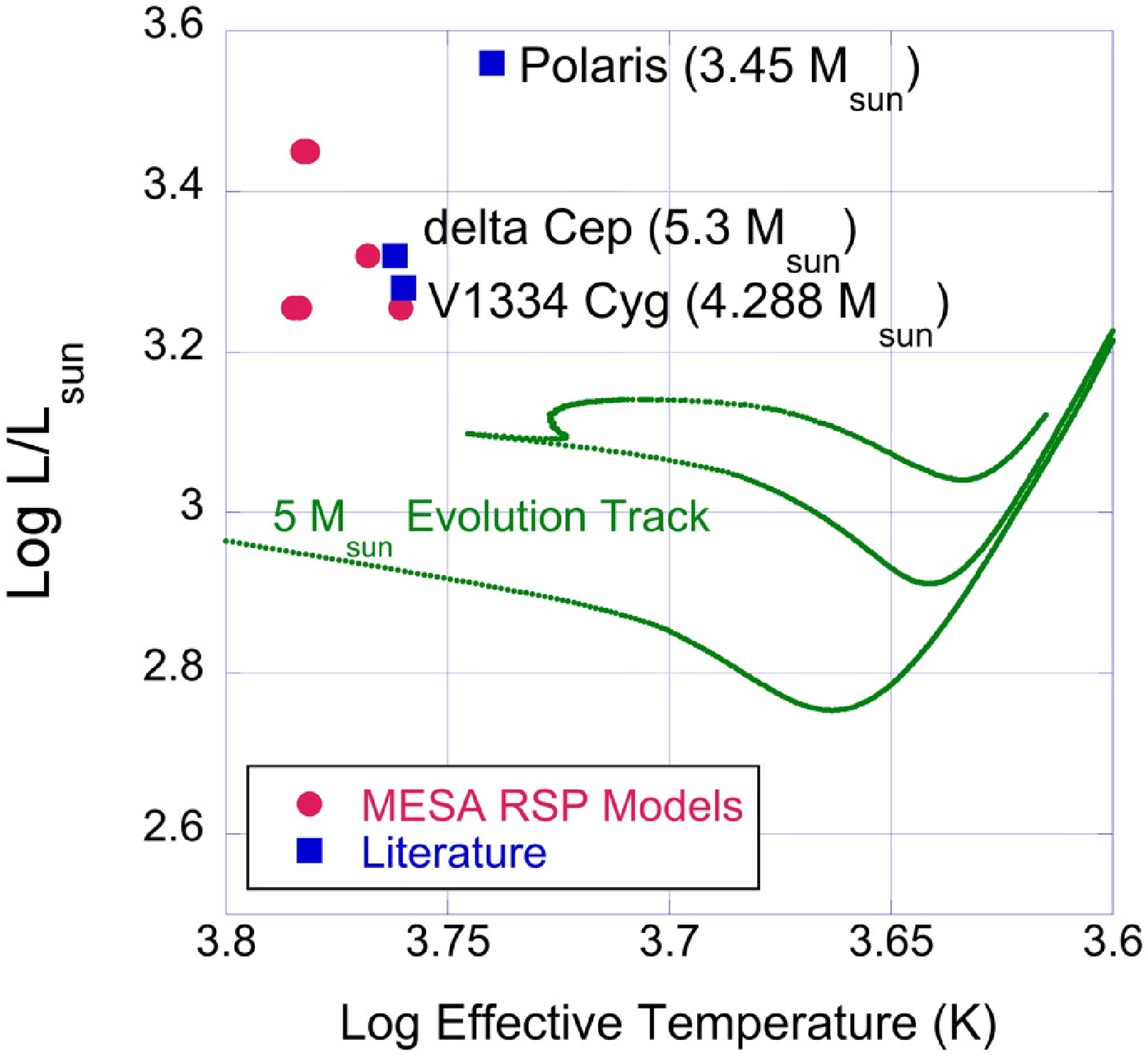}{HRD_3models}{H-R diagram showing locations of observed Cepheids, and of MESA RSP models that match the observed pulsation periods.  The green line shows a 5 M$_{\odot}$ MESA evolution track with (Y, Z) = (0.27, 0.015), without mass loss or rotation.}

\begin{table}[!ht]
\caption{Properties of MESA RSP models with standard opacities and with opacity enhancement (+$\kappa$)}
\label{table}
\begin{center}
{\small
\begin{tabular}{lccccccc}  
\tableline
\noalign{\smallskip}
& $\delta$ Cep &   & Polaris  & &  V1334 Cyg   &\\
\noalign{\smallskip}
\tableline
\noalign{\smallskip}
Property & Normal & +$\kappa$ &  Normal & +$\kappa$  & Normal & +$\kappa$ \\
\tableline 
Mass (M$_{\odot}$) &  5.3 & 5.3 & 5.93& 5.93 & 4.288 & 4.288 \\
Luminosity (L$_{\odot}$) & 2089 & 1800 & 2818 & 2818 & 1800 & 1800 \\
T$_{\rm eff}$ (K) & 5861 & 5761 & 6048 & 6064 & 6072 & 6092 \\
P$_{\rm o}$ (days) & 5.3676 & 5.3676 & 3.9716 & 3.9720 & 3.3322 & 3.3316 \\
Pulsation mode  & F & F  & 1st OT & 1st OT  & 1st OT & 1st OT \\
\noalign{\smallskip}
Linear Growth \\
Rate per Period & 0.00819 & 0.00119 & 0.00105 & 0.00268 & 0.00488 (F) & 0.00532 \\
  &      &      &     &    & 0.00688 (OT) &     \\
Limiting \\
Amplitude (km/s) & 22 & 7.3  & 3.6  & 0.61  & 18 (F)  & 11 (OT) \\
\tableline\
\end{tabular}
}
\end{center}
\end{table}

\vspace*{-1.0cm}

\section{Summary and Conclusions}
The MESA code and new RSP capabilities are useful tools to model Cepheid evolution and pulsation.  Increasing helium-burning reaction rates increases the luminosity and extent of Cepheid blue loops, but implausibly high increases are needed to resolve the mass discrepancy.  Including opacity enhancements around 200,000-400,000 K in the envelope, as proposed for SPB/$\beta$ Cep variables, does not significantly affect the evolution tracks or the extent of blue loops.  Radial velocity amplitudes of MESA RSP envelope models of $\delta$ Cep, Polaris, and V1334 Cyg agree well with observed amplitudes.  However, including an opacity enhancement as proposed for SPB/$\beta$ Cep stars results in lower radial velocity amplitudes, although these amplitudes could be increased by decreasing the viscous dissipation settings in the RSP models.  The Polaris and V1334 Cyg RSP models with periods matching observations have locations in the H-R diagram above that of a 5 M$_{\odot}$ evolution track, neglecting mass loss and rotation, inconsistent with their lower dynamical masses.



\acknowledgements This research was supported in part by the National Science Foundation under Grant No.~NSF ACI-1663688.  We are grateful to KITP for the opportunity to participate in the 2019 MESA summer school.  We  thank Bill Paxton, Frank Timmes, Josiah Schwab, the 2019 MESA summer school students, and LANL undergraduate summer student Stephanie Flynn.

\bibliography{Guzik_RRL2019v9}  

\end{document}